\documentclass[showpacs,preprint,preprintnumbers,nofootinbib,prd,eqsecnum,10pt]{revtex4-1}
  
\usepackage{amsmath,amssymb,amsfonts}
\usepackage{mathrsfs}
\usepackage[normalem]{ulem}
\usepackage{textcomp}
\usepackage[pdfusetitle]{hyperref}
\usepackage{bm}
\usepackage[bottom]{footmisc}
\usepackage{tensor}
\usepackage{graphicx}
\usepackage{psfrag}
\usepackage[usenames,dvipsnames]{xcolor}
\usepackage[utf8]{inputenc}
\usepackage[english, capitalise]{cleveref}

\crefname{chapter}{Chap.}{Chap.}
\crefname{section}{Sec.}{Sec.}
\Crefname{chapter}{Chapter}{Chapters}
\Crefname{section}{Section}{Sections}
\Crefname{eqs}{Eqs.}{Eqs.}
\Crefformat{eqs}{Eqs.~(#2#1#3)}

\newcommand{\dd}{\mathrm{d}}

\definecolor{darkgreen}{rgb}{0,0.5,0}

\hypersetup{
    bookmarks=true,
    unicode=false,
    pdftoolbar=true,
    pdfmenubar=true,
    pdffitwindow=false,
    pdfstartview={FitH},
    pdftitle={My title},
    pdfauthor={Author},
    pdfsubject={Subject},
    pdfcreator={Creator},
    pdfproducer={Producer},
    pdfkeywords={keyword1} {key2} {key3},
    pdfnewwindow=true,
    colorlinks=true,
    linkcolor=red,
    citecolor=cyan,
    filecolor=magenta,
    urlcolor=darkgreen,
    linktocpage=true
    pdftitle={Hamiltonian for tidal interactions in compact binary systems to next-to-next-to-leading post-Newtonian order},
    pdfsubject={Gravitation},
    pdfauthor={Quentin Henry, Guillaume Faye, Luc Blanchet},
    pdfkeywords={Post-Newtonian approximation}
  }

\allowdisplaybreaks

\begin{document}

\title{Hamiltonian for tidal interactions in compact binary systems to next-to-next-to-leading post-Newtonian order}

\author{Quentin \textsc{Henry}}\email{henry@iap.fr}
\affiliation{$\mathcal{G}\mathbb{R}\varepsilon{\mathbb{C}}\mathcal{O}$, 
Institut d'Astrophysique de Paris,\\ UMR 7095, CNRS, Sorbonne Universit{\'e},\\
98\textsuperscript{bis} boulevard Arago, 75014 Paris, France}

\author{Guillaume \textsc{Faye}}\email{faye@iap.fr}
\affiliation{$\mathcal{G}\mathbb{R}\varepsilon{\mathbb{C}}\mathcal{O}$, 
Institut d'Astrophysique de Paris,\\ UMR 7095, CNRS, Sorbonne Universit{\'e},\\
98\textsuperscript{bis} boulevard Arago, 75014 Paris, France}

\author{Luc \textsc{Blanchet}}\email{luc.blanchet@iap.fr}
\affiliation{$\mathcal{G}\mathbb{R}\varepsilon{\mathbb{C}}\mathcal{O}$, 
Institut d'Astrophysique de Paris,\\ UMR 7095, CNRS, Sorbonne Universit{\'e},\\
98\textsuperscript{bis} boulevard Arago, 75014 Paris, France}
 
\date{\today}

\begin{abstract}
In previous works, we obtained the leading, next-to-leading and next-to-next-to-leading (NNL) post-Newtonian (PN) corrections in the conservative tidal interactions between two compact non-spinning objects using a Lagrangian of effective field theory (EFT) in harmonic coordinates. In the present paper, we compute the corresponding NNL PN tidal effective Hamiltonian in ADM-like and isotropic coordinates, with contributions from mass quadrupole, current quadrupole and mass octupole tidal interactions, consistently included at that level. We also derive the NNL tidal Hamiltonian in Delaunay variables. We find full agreement in the overlap with recent results that were derived using tools from scattering amplitudes and the EFT to second post-Minkowskian (PM) order. 
\end{abstract}

\maketitle

\section{Introduction}\label{sec:introduction}

The first ever direct measurement of tidal interactions in compact binary systems were made through gravitational-wave observations~\cite{GW170817, Abbott_2020} (see~\cite{DHS20} for a recent review). Tidal interactions in compact binaries for which at least one of the companions is a neutron star depend on its internal equation of state (EoS). They start to affect the inspiral phase evolution at the dominant mass-quadrupole level $\mathcal{O}(\epsilon_\text{tidal})\sim (v/c)^{10}$, where $v$ is the relative velocity of the binary and $c$ the speed of light in vacuum, formally comparable to an orbital effect arising at the 5PN order.

Previous works on the problem of tidal interactions between neutron stars and their observability by gravitational-wave observatories include Refs.~\cite{FH08, Hind08, BinnP09, DN09tidal, HindLLR10, DNV12, BiniDF12, F14}. Recently, we have solved this problem up to the next-to-next-to-leading (NNL) order, \textit{i.e.} 2PN beyond the dominant order, which we refer to as $\mathcal{O}(\epsilon_\text{tidal}/c^4)\sim (v/c)^{14}$ (comparable to a 7PN orbital effect).

In Ref.~\cite{HFB20a}, we focused on the binary's equations of motion and conservative dynamics to NNL order. We used traditional PN methods~\cite{BFP98, BFeom} and a Lagrangian formulation based on the Fokker action with effective non-minimal matter couplings to gravity describing the compact bodies' internal structure. In our approach, to NNL order, we parametrize the tidal interactions by three tidal polarizability coefficients $\mu_{A}^{(2)}$, $\sigma_{A}^{(2)}$ and $\mu_{A}^{(3)}$ in the matter action, defined by (with $A=1,2$)~\cite{Hind08, HindLLR10}
\begin{align}\label{eq:defpolarizability}
G \mu_{A}^{(2)} = \frac{2}{3} \,k_{A}^{(2)}
R_{A}^{5}\,,\qquad G \sigma_{A}^{(2)} =
\frac{1}{48} \,j_{A}^{(2)} R_{A}^{5}\,,\qquad G \mu_{A}^{(3)} = \frac{2}{15} \,k_{A}^{(3)} R_{A}^{7}\,,
\end{align}
where $G$ is the gravitational constant, $R_A$ is the areal radius of the neutron star, and $k_{A}^{(2)}$, $j_{A}^{(2)}$ and $k_{A}^{(3)}$ are relativistic generalizations of Love numbers corresponding to the mass quadrupole, current quadrupole and mass octupole tidal interactions, respectively; we have $\mu^{(2)}=\mathcal{O}(\epsilon_\text{tidal})$, $\sigma^{(2)}=\mathcal{O}(\epsilon_\text{tidal})$, and $\mu^{(3)}=\mathcal{O}(\epsilon_\text{tidal}/c^4)$. The mass quadrupole tidal term yields the leading effect together with NL and NNL corrections, the current quadrupole term contains NL and NNL effects (because of an extra factor $1/c^2$ in the action), whereas the mass octupole term represents a pure NNL contribution.

Next, in Ref.~\cite{HFB20b}, we investigated the emitted gravitational radiation and flux to NNL order. We computed the stress-energy-tensor from the effective matter action and inserted it into a gravitational-wave generation formalism (namely, the MPM-PN formalism~\cite{BD86, B98mult}) to obtain the total energy flux at infinity and the time evolution of the gravitational-wave phase and frequency.    

In the present paper, we extend our first work~\cite{HFB20a} by deriving the Hamiltonian for (conservative) tidal effects to NNL order. Starting from the Lagrangian in harmonic coordinates, we perform some minimal shifts of the trajectories in order to remove accelerations (present in both the point-particle and tidal parts of the Lagrangian in harmonic coordinates) so as to construct an admissible Hamiltonian from an ordinary Legendre transformation. Then, we play with canonical transformations at the level of that Hamiltonian to define the dynamics in ADM-like coordinates (for just the point-particle part of the Hamiltonian) and in isotropic coordinates.

Recently, resorting to methods from scattering amplitudes and effective field theory (EFT)~\cite{GR06, Cheung:2018wkq, Bjerrum-Bohr:2018xdl}, the Hamiltonian for tidal effects, comprising the contributions from mass quadrupole and current quadrupole tidal moments, was obtained in the post-Minkowskian (PM) expansion (rather than the PN expansion) up to NL PM order, which means including the dominant $\sim G^2$ tidal terms and the next to dominant $\sim G^3$ ones~\cite{Cheung:2020sdj, Kalin:2020lmz}.\footnote{Previously, the amplitude/EFT program led to the derivation of the Hamiltonian for point particles (neglecting tidal effects) to 3PM order~\cite{Bern:2019nnu,klin2020postminkowskian,klin2020conservative}. This extended many previous works using more traditional methods deriving the Hamiltonian to 2PM order~\cite{WG79, WH80, Westpf85, BeDD81, LSB08, Dscatt16, Dscatt17, BFok18}.} In addition, the mass and current octupole contributions were computed at the same order in~\cite{Kalin:2020lmz}.

We check below that our results for the NL and NNL tidal terms in the PN approximation are in full consistency with the overlapping tidal terms of the PN expanded PM Hamiltonians in the literature~\cite{Cheung:2020sdj, Kalin:2020lmz}. While evidently our PN truncation does not allow us to control some high powers of $v/c$ provided by the latter PM works, we do obtain higher order PM terms $\sim G^4$ arising at the NNL PN $\mathcal{O}(\epsilon_\text{tidal}/c^4)$ level for the mass and current quadrupoles.

In Sec.~\ref{sec:lagrangian}, we recall from Ref.~\cite{HFB20a} the NNL tidal Lagrangian we begin with; we construct the corresponding Hamiltonians in Sec.~\ref{sec:hamiltonian} both in ADM-like and in isotropic coordinates, and in Sec.~\ref{sec:delaunay}, we derive the Delaunay form of the Hamiltonian, together with the invariant conservative energy and periastron advance for circular orbits.

\section{The tidal Lagrangian} \label{sec:lagrangian}

In Ref.~\cite{HFB20a}, the effective matter action is defined with specific non-minimal couplings that describe the tidal effects parametrized by the polarizability coefficients~\eqref{eq:defpolarizability}. After adding to the effective matter action the Einstein-Hilbert gravitational action (with standard harmonic-gauge fixing term), the associated Fokker action is constructed and the conservative dynamics and associated invariants are derived to NNL order.

For two compact bodies of masses $m_A$ in an arbitrary frame, the Lagrangian reads $L = L_\text{pp} + L_\text{tidal}$, where $L_\text{pp}$ stands for the standard point-particle (pp) part and $L_{\text{tidal}}$ for the tidal part. To be consistent with the NNL tidal part, we provide the well-known point-particle part in harmonic coordinates up to 2PN order~\cite{DD81b}:
\begin{align}\label{eq:Lpp}
L_\text{pp} &= \frac{m_1 v_1^2}{2} + \frac{G m_1 m_2}{2 r_{12}}
\nonumber \\ & 
%%%%%%%%%%%%%%%%%%%%%%%%%%%%%%%%%%%%%%%%%%%%%%%%%%%%%%%%%%%%%%%%%%%%%
+ \frac{1}{c^2} \Biggl[ - \frac{G^2 m_1^2 m_2}{2
	r_{12}^2} + \frac{m_1 v_1^4}{8} + \frac{G m_1 m_2}{r_{12}} \left(
- \frac{1}{4} (n_{12}v_1) (n_{12}v_2) + \frac{3}{2} v_1^2 -
\frac{7}{4} (v_1v_2) \right)  \Biggr] \nonumber \\ & 
%%%%%%%%%%%%%%%%%%%%%%%%%%%%%%%%%%%%%%%%%%%%%%%%%%%%%%%%%%%%%%%%%%%%%%
+ \frac{1}{c^4} \Biggl[ \frac{G^3 m_1^3 m_2}{2 r_{12}^3} + \frac{19
	G^3 m_1^2 m_2^2}{8 r_{12}^3} \nonumber \\ & \qquad ~\, + \frac{G^2
	m_1^2 m_2}{r_{12}^2} \left( \frac{7}{2} (n_{12}v_1)^2 - \frac{7}{2}
(n_{12}v_1) (n_{12}v_2) + \frac{1}{2}(n_{12}v_2)^2 + \frac{1}{4} v_1^2
- \frac{7}{4} (v_1v_2) + \frac{7}{4} v_2^2 \right) \nonumber \\ &
\qquad ~\, + \frac{G m_1 m_2}{r_{12}} \bigg( \frac{3}{16}
(n_{12}v_1)^2 (n_{12}v_2)^2 - \frac{7}{8} (n_{12}v_2)^2 v_1^2 +
\frac{7}{8} v_1^4 + \frac{3}{4} (n_{12}v_1) (n_{12}v_2) (v_1v_2)
\nonumber \\ & \qquad \qquad \qquad \qquad - 2 v_1^2 (v_1v_2) +
\frac{1}{8} (v_1v_2)^2 + \frac{15}{16} v_1^2 v_2^2 \bigg) + \frac{m_1
	v_1^6}{16} \nonumber \\ & \qquad ~\, + G m_1 m_2 \left( -
\frac{7}{4} (a_1 v_2) (n_{12}v_2) - \frac{1}{8} (n_{12} a_1)
(n_{12}v_2)^2 + \frac{7}{8} (n_{12} a_1) v_2^2 \right) \Biggr] +
1\leftrightarrow 2 + \mathcal{O}\left(\frac{1}{c^6}\right)\,.
\end{align}
Here, the particles' harmonic coordinate positions are denoted $\bm{y}_A=(y_A^i)$, together with their coordinate velocities and accelerations $v_A^i = \dd y_A^i/\dd t$ and $a_A^i = \dd v_A^i/\dd t$. We pose $r_{12}=\vert\bm{y}_1 - \bm{y}_2\vert$, $n_{12}^i=(y^i_1-y^i_2)/r_{12}$, and $(n_{12}v_1)=\bm{n}_{12}\cdot\bm{v}_1$, $(a_1 v_2)=\bm{a}_1\cdot\bm{v}_2$ and so on. Now, the tidal part is parametrized by the set of coefficients~\eqref{eq:defpolarizability} to NNL order, meaning that we neglect $\mathcal{O}(\epsilon_\text{tidal}/c^6)$ corrections. It is given in harmonic coordinates by~\cite{HFB20a}
\begin{align}\label{eq:Ltidal}
L_{\text{tidal}}&= \frac{G^2 m_{2}^2}{r_{12}^6} \Biggl\{\frac{3}{2} \mu_1^{(2)}
 \nonumber\\ &+ \frac{1}{c^{2}} \Biggl [\mu_1^{(2)} \biggl(- \frac{9}{2} (n_{12}{} v_{1}{})^2
 - 18 (n_{12}{} v_{1}{}) (n_{12}{} v_{2}{})
 + 18 (n_{12}{} v_{2}{})^2
 -  \frac{9}{2} (v_{1}{} v_{2}{})
 + \frac{15}{4} v_{1}{}^{2}\biggr)\nonumber\\
&\qquad \qquad + \sigma_1^{(2)} \biggl(-12 (n_{12}{} v_{12}{})^2
 + 12 v_{12}{}^{2}\biggr)
 -  \frac{3 G m_{1} \mu_1^{(2)}}{r_{12}}
 -  \frac{21 G m_{2} \mu_1^{(2)}}{2 r_{12}}\Biggl]
\nonumber\\ &+ \frac{1}{c^{4}} \Biggl [\mu_1^{(2)}
   \biggl(\frac{9}{2} (n_{12}{} v_{1}{})^4 - 18 (n_{12}{} v_{1}{})^3 (n_{12}{} v_{2}{})
 + 45 (n_{12}{} v_{1}{})^2 (n_{12}{} v_{2}{})^2
 - 54 (n_{12}{} v_{1}{}) (n_{12}{} v_{2}{})^3 + \frac{63}{2} (n_{12}{} v_{2}{})^4
 \nonumber\\
 &\qquad \qquad+ 9 (n_{12}{} v_{1}{}) (n_{12}{} v_{2}{}) (v_{1}{} v_{2}{}) - 18 (n_{12}{} v_{2}{})^2 (v_{1}{} v_{2}{})
 + \frac{9}{2} (v_{1}{} v_{2}{})^2
 - 9 (n_{12}{} v_{1}{})^2 v_{12}{}^{2}
+ 27 (n_{12}{} v_{1}{}) (n_{12}{} v_{2}{}) v_{12}{}^{2}
\nonumber\\
&\qquad \qquad - 36 (n_{12}{} v_{2}{})^2 v_{12}{}^{2} + 9 (v_{1}{} v_{2}{}) v_{12}{}^{2}
 + 9 v_{12}{}^{4}
 -  \frac{9}{4} (n_{12}{} v_{1}{})^2 v_{1}{}^{2}
 -  \frac{9}{2} (n_{12}{} v_{1}{}) (n_{12}{} v_{2}{}) v_{1}{}^{2}
 + \frac{27}{2} (n_{12}{} v_{2}{})^2 v_{1}{}^{2}
\nonumber\\
&\qquad \qquad - 9 (v_{1}{} v_{2}{}) v_{1}{}^{2} -  \frac{27}{4} v_{12}{}^{2} v_{1}{}^{2}
 + \frac{69}{16} v_{1}{}^{4}\biggr)
\nonumber\\
&\qquad + \mu_1^{(2)} r_{12} \biggl(-12 (v_{12}{} a_{2}{}) (n_{12}{} v_{1}{})
 + 60 (n_{12}{} a_{2}{}) (n_{12}{} v_{1}{})^2
 + 21 (v_{12}{} a_{2}{}) (n_{12}{} v_{2}{}) -  \frac{9}{2} (v_{1}{} a_{2}{}) (n_{12}{} v_{2}{})
\nonumber\\
&\qquad \qquad - 102 (n_{12}{} a_{2}{}) (n_{12}{} v_{1}{}) (n_{12}{} v_{2}{}) + 60 (n_{12}{} a_{2}{}) (n_{12}{} v_{2}{})^2
 + \frac{69}{2} (n_{12}{} a_{2}{}) (v_{1}{} v_{2}{})
 -  \frac{69}{4} (n_{12}{} a_{2}{}) v_{1}{}^{2} -  \frac{39}{2} (n_{12}{} a_{2}{}) v_{2}{}^{2}\biggr)
\nonumber\\
&\qquad + \sigma_1^{(2)} \biggl(60 (n_{12}{} v_{12}{})^4
 - 96 (n_{12}{} v_{12}{})^3 (n_{12}{} v_{1}{})
 + 48 (n_{12}{} v_{12}{})^2 (n_{12}{} v_{1}{})^2
 - 24 (n_{12}{} v_{12}{})^2 (v_{1}{} v_{2}{})\nonumber\\
&\qquad \qquad + 24 (n_{12}{} v_{12}{}) (n_{12}{} v_{1}{}) (v_{1}{} v_{2}{})
 + 12 (v_{1}{} v_{2}{})^2
 - 84 (n_{12}{} v_{12}{})^2 v_{12}{}^{2}
 + 96 (n_{12}{} v_{12}{}) (n_{12}{} v_{1}{}) v_{12}{}^{2}
 - 36 (n_{12}{} v_{1}{})^2 v_{12}{}^{2}\nonumber\\
&\qquad \qquad + 24 (v_{1}{} v_{2}{}) v_{12}{}^{2}
 + 24 v_{12}{}^{4}
 + 18 (n_{12}{} v_{12}{})^2 v_{1}{}^{2}
 - 24 (n_{12}{} v_{12}{}) (n_{12}{} v_{1}{}) v_{1}{}^{2}
 - 24 (v_{1}{} v_{2}{}) v_{1}{}^{2}
 - 18 v_{12}{}^{2} v_{1}{}^{2}
 + 12 v_{1}{}^{4}\biggr)\nonumber\\
&\qquad + \sigma_1^{(2)} r_{12} \biggl(16 (n_{12}{} a_{2}{}) (n_{12}{} v_{12}{})^2
 + 24 (v_{12}{} a_{2}{}) (n_{12}{} v_{1}{})
 - 24 (n_{12}{} a_{2}{}) (n_{12}{} v_{12}{}) (n_{12}{} v_{1}{})
 - 16 (n_{12}{} a_{2}{}) v_{12}{}^{2}\biggr)\nonumber\\
&\qquad + \frac{G m_{1} \mu_1^{(2)}}{r_{12}} \biggl(\frac{807}{8} (n_{12}{} v_{1}{})^2
 + \frac{381}{8} (n_{12}{} v_{1}{}) (n_{12}{} v_{2}{})
 - 138 (n_{12}{} v_{2}{})^2
 -  \frac{387}{8} (v_{1}{} v_{2}{})
 + \frac{63}{8} v_{1}{}^{2}
 + 42 v_{2}{}^{2}\biggr)\nonumber\\
&\qquad + \frac{G m_{2} \mu_1^{(2)}}{r_{12}} \biggl(\frac{27}{2} (n_{12}{} v_{1}{})^2
 + \frac{1051}{8} (n_{12}{} v_{1}{}) (n_{12}{} v_{2}{})
 -  \frac{865}{8} (n_{12}{} v_{2}{})^2
 + \frac{83}{8} (v_{1}{} v_{2}{})
 -  \frac{45}{4} v_{1}{}^{2}
 + \frac{49}{8} v_{2}{}^{2}\biggr)\nonumber\\
&\qquad + \frac{G m_{1} \sigma_1^{(2)}}{r_{12}} \biggl(-8 (n_{12}{} v_{12}{})^2
 + 8 v_{12}{}^{2}\biggr)
 + \frac{G m_{2} \sigma_1^{(2)}}{r_{12}} \biggl(36 (n_{12}{} v_{12}{})^2
 - 36 v_{12}{}^{2}\biggr)
\nonumber\\
&\qquad -  \frac{60 G^2 m_{1}^2 \mu_1^{(2)}}{7 r_{12}^2} + \frac{707 G^2 m_{1} m_{2} \mu_1^{(2)}}{8 r_{12}^2}
 + \frac{165 G^2 m_{2}^2 \mu_1^{(2)}}{4 r_{12}^2}\Biggl]
 + \frac{15 \mu_1^{(3)}}{2 r_{12}^2}\Biggl\}
 + 1 \leftrightarrow 2 + \mathcal{O}\left(
 \frac{\epsilon_\text{tidal}}{c^{6}}\right) \,.
\end{align}

In harmonic coordinates, the above Lagrangian is in fact a generalized one, which contains accelerations $\bm{a}_A = \dd \bm{v}_A/\dd t$ first arising at the 2PN order for the point-particle part and at the NNL/7PN order for the tidal part. Introducing the conjugate momenta $\bm{p}_A$ and $\bm{q}_A$ associated with the positions and velocities
\begin{equation}\label{eq:pqdef}
p_A^i \equiv \frac{\partial L}{\partial v_A^i} - \frac{\dd}{\dd t}\left(\frac{\partial L}{\partial a_A^i}\right)\,,\qquad
q_A^i \equiv \frac{\partial L}{\partial a_A^i} \,,
\end{equation}
the equations of motion and conservative energy are obtained from
\begin{equation}\label{eq:edmE}
\frac{\delta L}{\delta y_A^i} \equiv \frac{\partial L}{\partial y_A^i} - \frac{\dd p_A^i}{\dd t} = 0\,,\qquad E = \sum_{A=1,2} \Bigl(v_A^i p_A^i + a_A^i q_A^i\Bigr) - L\,.
\end{equation}
We can read, from~\eqref{eq:Lpp}--\eqref{eq:Ltidal}, the explicit dependence of the Lagrangian on accelerations, namely
\begin{subequations}\label{eq:qA}
\begin{align}
q^i_{1\,\text{pp}}={}&\frac{G m_{1} m_{2}}{c^4} \biggl [n_{12}^{i} \Bigl(-
 \frac{1}{4} (n_{12}{} v_{1}{}) (n_{12}{} v_{2}{})
 + \frac{1}{4} (v_{1}{} v_{2}{})\Bigr)
 + \frac{7}{4} (n_{12}{} v_{2}{}) v_{1}^{i}
 + \Bigl(\frac{1}{4} (n_{12}{} v_{1}{})
 - 2 (n_{12}{} v_{2}{})\Bigr) v_{2}^{i}\biggl] + \mathcal{O}\left(\frac{1}{c^6}\right)\,,\\
%%%%%%%%%%%%%%%%%%%%%%%%%%%%%%%%%%%%%%%%%%%%%%%%%%%%%%%%%%%%%%
q^i_{1\,\text{tidal}}={}&\frac{G^2 m_{1}^2}{c^4 r_{12}^5} \Biggl\{ n_{12}^{i}
\biggl [\Bigl(-60 \mu_2^{(2)}
- 16 \sigma_2^{(2)}\Bigr) (n_{12}{} v_{1}{})^2
+ \Bigl(102 \mu_2^{(2)}
+ 8 \sigma_2^{(2)}\Bigr) (n_{12}{} v_{1}{}) (n_{12}{} v_{2}{})
+ \Bigl(-60 \mu_2^{(2)}
+ 8 \sigma_2^{(2)}\Bigr) (n_{12}{} v_{2}{})^2\nonumber\\
&\qquad\qquad\quad + \Bigl(- \frac{69}{2} \mu_2^{(2)}
- 32 \sigma_2^{(2)}\Bigr) (v_{1}{} v_{2}{})
+ \Bigl(\frac{39}{2} \mu_2^{(2)}
+ 16 \sigma_2^{(2)}\Bigr) v_{1}{}^{2}
+ \Bigl(\frac{69}{4} \mu_2^{(2)}
+ 16 \sigma_2^{(2)}\Bigr) v_{2}{}^{2}\biggl]
\nonumber\\
&\qquad\quad + v_{1}^{i}\biggl [21 \mu_2^{(2)} (n_{12}{} v_{1}{}) + \Bigl(-12 \mu_2^{(2)}
+ 24 \sigma_2^{(2)}\Bigr) (n_{12}{} v_{2}{})\biggl] \nonumber\\
&\qquad\quad 
+ v_{2}^{i}\biggl [- \frac{33}{2} \mu_2^{(2)} (n_{12}{} v_{1}{})
+ \Bigl(12 \mu_2^{(2)}
- 24 \sigma_2^{(2)}\Bigr) (n_{12}{} v_{2}{})\biggl] \Biggl\} +  \mathcal{O}\left(\frac{\epsilon_\text{tidal}}{c^{6}}\right)\,.
\end{align}
\end{subequations}
 
\section{The tidal Hamiltonian} \label{sec:hamiltonian}

To construct the Hamiltonian, we need to remove the accelerations at 2PN and NNL/7PN orders by means of shifts of the particles' trajectories also known as ``contact'' transformations. For a generalized Lagrangian $L[\bm{y}_A, \bm{v}_A, \bm{a}_A]$ that is (i) linear in accelerations and such that (ii) the accelerations appear at the highest considered PN order (in our case 2PN and NNL/7PN order), the shifts read $\bm{y}_A\longrightarrow\bm{Y}_A = \bm{y}_A + \delta\bm{y}_A$ with~\cite{ABF01}
\begin{equation}\label{eq:deltayA}
\delta y_A^i = \frac{1}{m_A}\left( q_A^i + \frac{\partial F}{\partial v_A^i}\right) +  \mathcal{O}\left(\frac{1}{c^{6}},\frac{\epsilon_\text{tidal}}{c^{6}}\right)\,,
\end{equation}
where, in the present case, $q_A^i = q^i_{A\,\text{pp}} + q^i_{A\,\text{tidal}}$ is the conjugate momentum~\eqref{eq:qA}, while $F = F_{\text{pp}} + F_{\text{tidal}}$ is an arbitrary function of the positions and velocities present at the highest 2PN and NNL/7PN levels. The effect of this contact transformation, combined with the addition of a total time derivative, yields the physically equivalent Lagrangian $L\longrightarrow L + \delta L$ with the extra contribution
\begin{equation}\label{eq:L'}
\delta L = \sum_{A=1,2} \frac{\delta L}{\delta y_A^i} \delta y_A^i + \frac{\dd F}{\dd t} + \mathcal{O}\left(\frac{1}{c^{6}},\frac{\epsilon_\text{tidal}}{c^{6}}\right)\,.
\end{equation}
Here, $\delta L/\delta y_A^i$ denotes the functional derivative of the Lagrangian as defined in Eq.~\eqref{eq:edmE} but evaluated \textit{off-shell}, without replacement of the accelerations. With the choice~\eqref{eq:deltayA}, the new Lagrangian is now ordinary, \textit{i.e.}, depends only on positions and velocities.

As we already said, the function $F$ can be adjusted at will and, for the point-particle case, it can be chosen in such a way that, starting from the harmonic-coordinates Lagrangian, the target Lagrangian uses position variables corresponding to ADM coordinates. For convenience, we adopt for $F$ the same point-particle part as in ADM coordinates, without tidal terms and up to 2PN order. We reserve our freedom of choosing the target coordinate system by adjusting the canonical transformations at the level of the Hamiltonian. Thus, according to Eq.~(4.15) in~\cite{ABF01}, we take:
\begin{equation}\label{eq:F}
F = \frac{G m_1 m_2}{c^4} \Bigg\{ \frac{G m_1}{r_{12}} \bigg( \frac{7}{4}
(n_{12}v_1) - \frac{1}{4} (n_{12}v_2) \bigg) + \frac{1}{4} (n_{12}v_2) v_1^2 \Bigg\} + 1 \leftrightarrow 2 + \mathcal{O}\left(\frac{1}{c^{6}}\right)\,.
\end{equation}

After the specific contact transformation~\eqref{eq:deltayA}--\eqref{eq:F}, the Lagrangian $L'=L+\delta L$ is a functional of trajectories $Y^i_A=y^i_A+\delta y^i_A$ and velocities $V^i_A=\dd Y^i_A/\dd t$. The Hamiltonian $H'$ follows from the usual Legendre transformation. Denoting by $\bm{P}_A$ the conjugate momenta $P^i_A=\partial L'/\partial V^i_A$, and posing also $R_{12}=\vert\bm{Y}_1 - \bm{Y}_2\vert$, $N^i_{12}=(Y^i_1-Y^i_2)/R_{12}$, $(N_{12}P_1)=\bm{N}_{12}\cdot\bm{P}_1$ \textit{etc.}, we find $H' = H'_\text{pp} + H'_\text{tidal}$, where the point-particle part reproduces the Hamiltonian in ADM coordinates to the considered order~\cite{JaraS15}:
\begin{align}\label{eq:H'pp}
H'_\text{pp} &= \frac{P_1^2}{2 m_1} - \frac{G m_1 m_2}{2 R_{12}}
\nonumber \\ & 
%%%%%%%%%%%%%%%%%%%%%%%%%%%%%%%%%%%%%%%%%%%%%%%%%%%%%%%%%%%%%%%%%%%%%%%
+ \frac{1}{c^2} \left\{  - \frac{P_1^4}{8 m_1^3} +
\frac{G^2 m_1^2 m_2}{2 R_{12}^2}
+ \frac{G m_1 m_2}{R_{12}} \left( \frac{1}{4}
\frac{(N_{12}P_1) (N_{12}P_2)}{ m_1 m_2} - \frac{3}{2}
\frac{P_1^2}{m_1^2} + \frac{7}{4} \frac{(P_1P_2)}{m_1 m_2} \right) \right\}
\nonumber \\& 
%%%%%%%%%%%%%%%%%%%%%%%%%%%%%%%%%%%%%%%%%%%%%%%%%%%%%%%%%%%%%%%%%%%%%%%
+ \frac{1}{c^4} \Bigg\{  \frac{P_1^6}{16 m_1^5} -
\frac{G^3 m_1^3 m_2}{4 R_{12}^3} - \frac{5 G^3 m_1^2 m_2^2}{8R_{12}^3}
\nonumber \\
& \qquad ~
+ \frac{G^2 m_1^2 m_2}{R_{12}^2} \left(  - \frac{3}{2}
\frac{(N_{12}P_1) (N_{12}P_2)}{m_1 m_2} + \frac{19}{4}
\frac{P_1^2}{m_1^2} - \frac{27}{4} \frac{(P_1P_2)}{m_1 m_2} +
\frac{5 P_2^2}{2 m_2^2} \right)
\nonumber \\
& \qquad ~
+ \frac{G m_1 m_2}{R_{12}} \bigg(  - \frac{3}{16}
\frac{(N_{12}P_1)^2 (N_{12}P_2)^2}{m_1^2 m_2^2} + \frac{5}{8}
\frac{(N_{12}P_2)^2 P_1^2}{m_1^2 m_2^2}
\nonumber \\
& \qquad \qquad 
+ \frac{5}{8} \frac{P_1^4}{m_1^4} - \frac{3}{4}
\frac{(N_{12}P_1) (N_{12}P_2) (P_1P_2)}{m_1^2 m_2^2}
- \frac{1}{8} \frac{(P_1P_2)^2}{m_1^2 m_2^2} - \frac{11}{16}
\frac{P_1^2 P_2^2}{m_1^2 m_2^2} \bigg) \Bigg\} + 1\leftrightarrow 2 + \mathcal{O}\left(\frac{1}{c^{6}}\right)\,,
\end{align}
and where the tidal part, accurate up to NNL/7PN order, reads
\begin{align}\label{eq:H'tidal}
H'_\text{tidal} &= \frac{G^2 m_2^2}{R_{12}^6}\Bigg\{- \frac{3}{2} \mu_1^{(2)}\nonumber\\
&\quad + \frac{1}{c^2} \Biggl [- \frac{12 \sigma_1^{(2)} P_{2}{}^{2}}{m_{2}^2}
+ \frac{(N_{12} P_{2})^2}{m_{2}^2} \Bigl(-18 \mu_1^{(2)}
+ 12 \sigma_1^{(2)}\Bigr)
+ \frac{(N_{12} P_{1}) (N_{12} P_{2})}{m_{1} m_{2}} \Bigl(18 \mu_1^{(2)}
- 24 \sigma_1^{(2)}\Bigr)\nonumber\\
&\qquad\quad + \frac{(P_{1} P_{2})}{m_{1} m_{2}} \Bigl(\frac{9}{2} \mu_1^{(2)}
+ 24 \sigma_1^{(2)}\Bigr)
+ \frac{(N_{12} P_{1})^2}{m_{1}^2} \Bigl(\frac{9}{2} \mu_1^{(2)}
+ 12 \sigma_1^{(2)}\Bigr)
+ \frac{P_{1}{}^{2}}{m_{1}^2} \Bigl(- \frac{15}{4} \mu_1^{(2)}
- 12 \sigma_1^{(2)}\Bigr)\nonumber\\
&\qquad\quad  + \frac{G}{R_{12}} \Bigl(3 m_{1} \mu_1^{(2)} + \frac{21}{2} m_{2} \mu_1^{(2)}\Bigr)\Biggl]\nonumber\\
&\quad + \frac{1}{c^4} \Biggl[\frac{(N_{12} P_{2})^4}{m_{2}^4} \Bigl(- \frac{63}{2} \mu_1^{(2)}
- 60 \sigma_1^{(2)}\Bigr)
+ \frac{P_{2}{}^{4}}{m_{2}^4} \Bigl(-9 \mu_1^{(2)}
- 12 \sigma_1^{(2)}\Bigr)
+ \frac{(P_{1} P_{2}) P_{2}{}^{2}}{m_{1} m_{2}^3} \Bigl(\frac{99}{4}
\mu_1^{(2)} + 60 \sigma_1^{(2)}\Bigr)
\nonumber\\
&\qquad\quad + \frac{(N_{12} P_{2})^2}{m_{2}^2} \biggl [\frac{P_{2}{}^{2}}{m_{2}^2} \Bigl(54 \mu_1^{(2)}
+ 72 \sigma_1^{(2)}\Bigr)
+ \frac{(P_{1} P_{2})}{m_{1} m_{2}} \Bigl(-54 \mu_1^{(2)}
- 144 \sigma_1^{(2)}\Bigr)\biggl]\nonumber\\
&\qquad\quad  + \frac{(P_{1} P_{2})^2}{m_{1}^2 m_{2}^2} \Bigl(- \frac{45}{2} \mu_1^{(2)} - 60 \sigma_1^{(2)}\Bigr)
+ \frac{(N_{12} P_{1})^3 (N_{12} P_{2})}{m_{1}^3 m_{2}} \Bigl(18 \mu_1^{(2)}
+ 48 \sigma_1^{(2)}\Bigr)
\nonumber\\
&\qquad\quad + \frac{P_{1}{}^{2}}{m_{1}^2} \biggl [\frac{(N_{12} P_{2})^2}{m_{2}^2} \Bigl(\frac{45}{2} \mu_1^{(2)}
+ 66 \sigma_1^{(2)}\Bigr) + \frac{P_{2}{}^{2}}{m_{2}^2} \Bigl(- \frac{45}{4} \mu_1^{(2)} - 30 \sigma_1^{(2)}\Bigr)
+ \frac{(P_{1} P_{2})}{m_{1} m_{2}} \Bigl(\frac{81}{4} \mu_1^{(2)}
+ 48 \sigma_1^{(2)}\Bigr)\biggl]
\nonumber\\
&\qquad\quad + \frac{(N_{12} P_{1})}{m_{1}} \biggl(\frac{(N_{12} P_{2})^3}{m_{2}^3} \Bigl(54 \mu_1^{(2)}
+ 144 \sigma_1^{(2)}\Bigr) + \frac{(N_{12} P_{2}) P_{1}{}^{2}}{m_{1}^2 m_{2}} \Bigl(- \frac{63}{2}
\mu_1^{(2)} - 48 \sigma_1^{(2)}\Bigr)\nonumber\\
&\qquad\qquad\quad + \frac{(N_{12} P_{2})}{m_{2}} \biggl [\frac{P_{2}{}^{2}}{m_{2}^2} \Bigl(-36 \mu_1^{(2)}
- 60 \sigma_1^{(2)}\Bigr)
+ \frac{(P_{1} P_{2})}{m_{1} m_{2}} \Bigl(45 \mu_1^{(2)}
+ 120 \sigma_1^{(2)}\Bigr)\biggl]\biggl)\nonumber\\
&\qquad\quad
+ \frac{(N_{12} P_{1})^4}{m_{1}^4} \Bigl(- \frac{9}{2} \mu_1^{(2)}
- 12 \sigma_1^{(2)}\Bigr)
+ \frac{P_{1}{}^{4}}{m_{1}^4} \Bigl(- \frac{45}{16} \mu_1^{(2)}
- 6 \sigma_1^{(2)}\Bigr)\nonumber\\
&\qquad\quad 
+ \frac{(N_{12} P_{1})^2}{m_{1}^2} \biggl [\frac{(N_{12} P_{2})^2}{m_{2}^2}
\Bigl(-45 \mu_1^{(2)} - 120 \sigma_1^{(2)}\Bigr)
+ \frac{P_{2}{}^{2}}{m_{2}^2} \Bigl(9 \mu_1^{(2)}
+ 24 \sigma_1^{(2)}\Bigr)\nonumber\\
&\qquad\qquad\quad 
+ \frac{(P_{1} P_{2})}{m_{1} m_{2}} \Bigl(-18 \mu_1^{(2)}
- 48 \sigma_1^{(2)}\Bigr) + \frac{P_{1}{}^{2}}{m_{1}^2} \Bigl(\frac{27}{4} \mu_1^{(2)}
+ 18 \sigma_1^{(2)}\Bigr)\biggl] \nonumber\\
&\qquad\quad+ \frac{G}{R_{12}} \biggl(m_{1} \biggl [\frac{(N_{12} P_{2})^2}{m_{2}^2} \Bigl(207 \mu_1^{(2)}
- 80 \sigma_1^{(2)}\Bigr)
+ \frac{P_{2}{}^{2}}{m_{2}^2} \Bigl(- \frac{45}{2} \mu_1^{(2)}
+ 80 \sigma_1^{(2)}\Bigr)\nonumber\\
&\qquad\qquad\quad\quad
+ \frac{(N_{12} P_{1}) (N_{12} P_{2})}{m_{1} m_{2}} \Bigl(- \frac{1341}{8}
\mu_1^{(2)} + 172 \sigma_1^{(2)}\Bigr)
+ \frac{(P_{1} P_{2})}{m_{1} m_{2}} \Bigl(\frac{3}{8} \mu_1^{(2)}
- 172 \sigma_1^{(2)}\Bigr)
\nonumber\\
&\qquad\qquad\quad\quad+ \frac{(N_{12} P_{1})^2}{m_{1}^2} \Bigl(- \frac{183}{2} \mu_1^{(2)}
- 92 \sigma_1^{(2)}\Bigr) + \frac{P_{1}{}^{2}}{m_{1}^2} \Bigl(\frac{123}{4} \mu_1^{(2)}
+ 92 \sigma_1^{(2)}\Bigr)\biggl] \nonumber\\
&\qquad\qquad\quad  + m_{2} \biggl [\frac{(N_{12} P_{2})^2}{m_{2}^2} \Bigl(\frac{331}{2} \mu_1^{(2)}
- 120 \sigma_1^{(2)}\Bigr)
+ \frac{P_{2}{}^{2}}{m_{2}^2} \Bigl(\frac{61}{4} \mu_1^{(2)}
+ 120 \sigma_1^{(2)}\Bigr)\nonumber\\
& \qquad\qquad\quad\quad
+ \frac{(N_{12} P_{1}) (N_{12} P_{2})}{m_{1} m_{2}} \Bigl(- \frac{1189}{8}
\mu_1^{(2)} + 228 \sigma_1^{(2)}\Bigr)
+ \frac{(P_{1} P_{2})}{m_{1} m_{2}} \Bigl(- \frac{401}{8} \mu_1^{(2)}
- 228 \sigma_1^{(2)}\Bigr)\nonumber\\
&\qquad\qquad\quad\quad  + \frac{(N_{12} P_{1})^2}{m_{1}^2} \Bigl(- \frac{81}{2} \mu_1^{(2)}
- 108 \sigma_1^{(2)}\Bigr)
+ \frac{P_{1}{}^{2}}{m_{1}^2} \Bigl(\frac{135}{4} \mu_1^{(2)} + 108
\sigma_1^{(2)}\Bigr)\biggl]\biggl)\nonumber
\\ &\qquad\quad
+ \frac{G^2}{R_{12}^2} \Bigl(\frac{303}{28} m_{1}^2 \mu_1^{(2)}
-  \frac{455}{8} m_{1} m_{2} \mu_1^{(2)}
- 39 m_{2}^2 \mu_1^{(2)}\Bigr)\Biggl]
-  \frac{15 \mu_1^{(3)}}{2 R_{12}^2} \Bigg\} + 1\leftrightarrow 2 +  \mathcal{O}\left(\frac{\epsilon_\text{tidal}}{c^{6}}\right)\,.
\end{align}

From now on, we restrict attention to the frame of the center of mass (CoM), for which the relative canonical momentum is simply $\bm{P} \equiv \bm{P}_1 = -\bm{P}_2$. Setting $R=R_{12}$ and $\bm{N}=\bm{N}_{12}$, we further change notation to introduce appropriate reduced variables 
\begin{equation}
\hat{H}' = \frac{H'}{m\nu}\,,\qquad \hat{\bm{P}} = \frac{\bm{P}}{m\nu}\,,\qquad \hat{R} = \frac{R}{G m}\,,
\end{equation}
where $m=m_1+m_2$ is the total mass, $\nu=m_1 m_2/m^2$ the symmetric mass ratio, and we will use later $\Delta=(m_1-m_2)/m$. In the CoM frame, it is also convenient to redefine the polarizability coefficients~\eqref{eq:defpolarizability} as~\cite{HFB20a, HFB20b}
\begin{equation}\label{eq:polarpm}
\widetilde{\mu}_\pm^{(\ell)} = \frac{G}{2}\left(\frac{c^2}{G m}\right)^{2\ell+1} \left(\frac{m_{2}}{m_{1}}\,\mu_{1}^{(\ell)} \pm \frac{m_{1}}{m_{2}}\,\mu_{2}^{(\ell)}\right)\,,\qquad \widetilde{\sigma}_\pm^{(\ell)} =
\frac{G}{2}\left(\frac{c^2}{G m}\right)^{2\ell+1} \left(\frac{m_{2}}{m_{1}}\,\sigma_{1}^{(\ell)} \pm
\frac{m_{1}}{m_{2}}\,\sigma_{2}^{(\ell)}\right)\,.
\end{equation}

With the Hamiltonian~\eqref{eq:H'pp}--\eqref{eq:H'tidal} in hand, we have a large freedom of variables provided by arbitrary canonical transformations. On the other hand, the community of scattering amplitudes and the EFT are deriving Hamiltonians in the PM approximation using isotropic coordinates, say $(\bm{\rho}, \bm{p})$~\cite{Bern:2019nnu}. Isotropic coordinates drastically simplify the expression of the Hamiltonian, which then depends on the momentum $\bm{p}$ only through the norm $\bm{p}^2\equiv p^2$ and not on the radial component $\bm{p}\cdot\bm{n}$ separately (with $\bm{n}=\bm{\rho}/\rho$). Thus, we now perform a canonical transformation from the reduced variables $(\hat{\bm{X}}, \hat{\bm{P}})$ to new (reduced versions of the) isotropic variables $(\hat{\bm{\rho}}, \hat{\bm{p}})$. We conveniently choose the generating function $\hat{G}(\hat{\bm{X}}, \hat{\bm{p}})$ of this transformation to be  
\begin{subequations}
\begin{align}
\hat{G}_{\text{pp}}=(\hat{N} \hat{p})\Bigg\{{}&\hat{R}
+ \frac{\nu}{2 c^2}
+ \frac{1}{c^4} \biggl [\frac{1}{\hat{R}} \Bigl(- \frac{1}{4} \nu
+ \frac{1}{4} \nu^2\Bigr)
+ \frac{1}{8} \nu^2 (\hat{N} \hat{p})^2
+ \Bigl(\frac{1}{4} \nu
-  \frac{1}{8} \nu^2\Bigr) \hat{p}^{2}\biggl]\Bigg\} + \mathcal{O}\left(\frac{1}{c^{6}}\right) \,,\\
%%%%%%%%%%%%%%%%%%%%%%%%%%%%%%%%%%%%%%%%%%%%%%%%%%%%%%%%%%%%
\hat{G}_{\text{tidal}}=(\hat{N} \hat{p})\Bigg\{{}&\frac{1}{c^{12} \hat{R}^5}
\biggl[ \tilde{\mu}_{+}^{(2)} \Bigl(\frac{9}{4} 
+ \frac{3}{2} \nu \Bigr)
+ \frac{15}{4} \tilde{\mu}_{-}^{(2)} \Delta
- 4 \tilde{\sigma}_{+}^{(2)}\biggl]
+ \frac{1}{c^{14}} \Biggl[\frac{1}{\hat{R}^6} \biggl [\tilde{\mu}_{+}^{(2)} \Bigl(-18
+ \frac{71}{4} \nu
-  \frac{9}{4} \nu^2\Bigr)\nonumber\\
& 
+ \tilde{\mu}_{-}^{(2)} \Delta \Bigl(- \frac{363}{14}-  \frac{15}{2} \nu \Bigr)
+ \tilde{\sigma}_{+}^{(2)} \Bigl(\frac{114}{7}
+ 16 \nu \Bigr)
-  \frac{34}{7} \tilde{\sigma}_{-}^{(2)} \Delta\biggl]
+ \frac{1}{\hat{R}^5} \biggl(\biggl [\tilde{\mu}_{+}^{(2)} \Bigl(\frac{9}{2}
- 9 \nu
+ \frac{9}{4} \nu^2\Bigr)
\nonumber\\
&  + \tilde{\mu}_{-}^{(2)} \Delta \Bigl(\frac{27}{8}
-  \frac{9}{4} \nu \Bigr)
+ \tilde{\sigma}_{+}^{(2)} \Bigl(9- 12 \nu \Bigr)
+ 6 \tilde{\sigma}_{-}^{(2)} \Delta\biggl] (\hat{N} \hat{p})^2
+ \biggl [\tilde{\mu}_{+}^{(2)} \Bigl(- \frac{27}{4}
+ \frac{81}{8} \nu
-  \frac{3}{8} \nu^2\Bigr)
\nonumber\\
&  + \tilde{\mu}_{-}^{(2)} \Delta \Bigl(- \frac{69}{16}
+ \frac{9}{4} \nu \Bigr)
+ \tilde{\sigma}_{+}^{(2)} \Bigl(- \frac{25}{2}
+ 10 \nu \Bigr)- 6 \tilde{\sigma}_{-}^{(2)} \Delta\biggl]
\hat{p}^{2}\biggl)\Biggl]\Bigg\} +  \mathcal{O}\left(\frac{\epsilon_\text{tidal}}{c^{6}}\right)\,,
\end{align}
\end{subequations}
with the canonical transformation being specified by $\hat{\bm{P}}=\partial\hat{G}/\partial\hat{\bm{X}}$ and $\hat{\bm{\rho}}=\partial\hat{G}/\partial\hat{\bm{p}}$. The Hamiltonian in isotropic coordinates is obtained as $\hat{H}^{\text{iso}}(\hat{\bm{\rho}}, \hat{\bm{p}}) = \hat{H}'(\hat{\bm{X}}, \hat{\bm{P}})$. To NNL order, we get
\begin{subequations}\label{eq:Hiso}
\begin{align}
\hat{H}_{\text{pp}}^{\text{iso}}={}&- \frac{1}{\hat{\rho}}
+ \frac{1}{2} \hat{p}^{2}\\& + \frac{1}{c^2} \biggl [\frac{1}{\hat{\rho}^2} \Bigl(\frac{1}{2}
+ \frac{1}{2} \nu \Bigr)
+ \frac{\hat{p}^{2}}{\hat{\rho}} \Bigl(- \frac{3}{2}
-  \nu \Bigr)
+ \Bigl(- \frac{1}{8}
+ \frac{3}{8} \nu \Bigr) \hat{p}^{4}\biggl]
\nonumber\\& + \frac{1}{c^4} \biggl [\frac{1}{\hat{\rho}^3} \Bigl(- \frac{1}{4}
-  \frac{3}{2} \nu \Bigr) + \frac{\hat{p}^{2}}{\hat{\rho}^2} \Bigl(\frac{5}{2}
+ \frac{27}{4} \nu+ \frac{3}{4} \nu^2\Bigr)
+ \frac{\hat{p}^{4}}{\hat{\rho}} \Bigl(\frac{5}{8}
-  \frac{5}{2} \nu
-  \nu^2\Bigr)
+ \Bigl(\frac{1}{16}
-  \frac{5}{16} \nu
+ \frac{5}{16} \nu^2\Bigr) \hat{p}^{6}\biggl] + \mathcal{O}\left(\frac{1}{c^{6}}\right)\,,\nonumber\\
%%%%%%%%%%%%%%%%%%%%%%%%%%%%%%%%%%%%%%%%%%%%%%%%%%%%%%%%%%%%%%
\hat{H}_{\text{tidal}}^{\text{iso}}={}&- \frac{3 \tilde{\mu}_{+}^{(2)}}{c^{10} \hat{\rho}^6}
\nonumber\\& + \frac{1}{c^{12}} \biggl(\frac{\hat{p}^{2}}{\hat{\rho}^6} \biggl
[\tilde{\mu}_{+}^{(2)} \Bigl(-6 
- 3 \nu \Bigr)
- 20 \tilde{\sigma}_{+}^{(2)}\biggl]
+ \frac{1}{\hat{\rho}^7} \biggl [\tilde{\mu}_{+}^{(2)} \Bigl(\frac{63}{4}
+ \frac{21}{2} \nu \Bigr)
-  \frac{15}{4} \tilde{\mu}_{-}^{(2)} \Delta
- 4 \tilde{\sigma}_{+}^{(2)}\biggl]\biggl)\nonumber\\
& + \frac{1}{c^{14}} \biggl(\frac{\hat{p}^{4}}{\hat{\rho}^6} \biggl
[\tilde{\mu}_{+}^{(2)} \Bigl(- \frac{63}{16}
-  \frac{21}{2} \nu
- 3 \nu^2\Bigr)
+ \tilde{\sigma}_{+}^{(2)} \Bigl(- \frac{15}{2}
- 40 \nu \Bigr)\biggl]
+ \frac{\hat{p}^{2}}{\hat{\rho}^7} \biggl [\tilde{\mu}_{+}^{(2)} \Bigl(\frac{261}{8}
+ \frac{699}{8} \nu
+ \frac{81}{8} \nu^2\Bigr)\nonumber\\
&\qquad + \tilde{\mu}_{-}^{(2)} \Delta \Bigl(- \frac{1989}{112}
-  \frac{15}{4} \nu \Bigr)
+ \tilde{\sigma}_{+}^{(2)} \Bigl(\frac{1977}{14}
+ 126 \nu \Bigr)
-  \frac{204}{7} \tilde{\sigma}_{-}^{(2)} \Delta\biggl]
+ \frac{1}{\hat{\rho}^8} \biggl [\tilde{\mu}_{+}^{(2)} \Bigl(- \frac{339}{7}
-  \frac{666}{7} \nu
- 15 \nu^2\Bigr)\nonumber\\
&\qquad + \tilde{\mu}_{-}^{(2)} \Delta \Bigl(\frac{141}{7}
+ 15 \nu \Bigr)
+ \tilde{\sigma}_{+}^{(2)} \Bigl(\frac{142}{7}
+ 20 \nu \Bigr)
-  \frac{34}{7} \tilde{\sigma}_{-}^{(2)} \Delta
- 15 \tilde{\mu}_{+}^{(3)}\biggl]\biggl) +  \mathcal{O}\left(\frac{\epsilon_\text{tidal}}{c^{6}}\right)\,.\label{eq:Hisotidal}
\end{align}
\end{subequations}
This Hamiltonian can be compared to the PN expansion of the Hamiltonian derived by the EFT/amplitude community. Of course, we recover for $\hat{H}^\mathrm{iso}_\text{pp}$ the 2PN expansion of the 3PM Hamiltonian given in Eq.~(10.8) of~\cite{Bern:2019crd}.\footnote{For point-particles without internal structure, the $n$PM Hamiltonian permits controlling the $(n-1)$PN approximation. Thus, the 3PM conservative Hamiltonian is sufficient to completely control the 2PN conservative dynamics.} Gladly, we also find a complete agreement for $\hat{H}_{\text{tidal}}^{\text{iso}}$ with the PN expansion of the NL PM tidal Hamiltonians in Eq.~(7) of~\cite{Cheung:2020sdj} and Eqs.~(19-20) of~\cite{Kalin:2020lmz}. Namely, we agree with the overlapping terms of the mass and current quadrupoles up to order $G^3$ and up to the NL/6PN order $\mathcal{O}(\epsilon_\text{tidal}/c^2)$. We also agree with the leading $G^2$ order of the mass octupole in~\cite{Kalin:2020lmz} up to NL/6PN order. Note that the NL PM approximation computed in~\cite{Cheung:2020sdj, Kalin:2020lmz} gives all the PN tidal terms at orders $G^2$ and $G^3$ up to NL/6PN order but overlooks those in $G^4$ arising at NNL/7PN [see the last line of Eq.~\eqref{eq:H'tidal}].

\section{Delaunay Hamiltonian and dynamical invariants} \label{sec:delaunay}

We provide in this section the Delaunay form of the Hamiltonian in the CoM frame. Starting from the canonical variables $(\bm{\rho}, \bm{p})$, and the associated ordinary velocity $v^i=\dd\rho^i/\dd t$, we parametrize $(\bm{\rho}, \bm{v})$ by means of usual osculating elliptic elements $(a, e, I, \ell, g, \Omega)$, namely the semi-major axis $a$, the eccentricity $e$, the inclination $I$, the argument of periastron $g$, the mean anomaly $\ell=n(t-t_0)$ where $n=(G m/a^3)^{1/2}$ is the mean motion, with $t_0$ denoting the instant of passage at periastron, such that the period between successive passages is given by $P=2\pi/n$, and the longitude of the ascending node $\Omega$. Then, the elliptical Delaunay variables are defined by $(\ell, g, \Omega; \Lambda, J, K)$ with $\Lambda=\mu(G m a)^{1/2}$, $J=\Lambda(1-e^2)^{1/2}$ (which is the orbital angular momentum) and $K=J \cos I$. The point is that the Delaunay variables are canonical, with $\ell, g, \Omega$ being the generalized positions and $\Lambda, J, K$ the generalized momenta (see \textit{e.g.}~\cite{laskar2017andoyer}). 

It is enough to consider the restricted planar problem,\footnote{The tidal fields only depend on the positions and velocities of the point particles, so does the dynamics, as is clear from the Lagrangian~\eqref{eq:Ltidal}. Therefore, in the CoM frame, the configuration variables are just $\boldsymbol{x}=\boldsymbol{y}_1-\boldsymbol{y}_2$ and $\boldsymbol{v}=\boldsymbol{v}_1-\boldsymbol{v}_2$. This implies that any axial vector constructed from those, in the absence of spin or intrinsic body deformations, is necessarily proportional to $\boldsymbol{x}\times \boldsymbol{v}$. In particular, this shows that the conserved angular momentum is orthogonal to both $\boldsymbol{x}$ and $\boldsymbol{v}$, so we conclude that the motion is planar.} in which case they reduce to $(\ell, g; \Lambda, J)$. The action variable $\Lambda$ is closely related to the radial action (with $r_\text{p}$ and $r_\text{a}$ denoting the apastron and periastron radii):
\begin{equation} 
I_r\equiv \frac{2}{2\pi} \int_{r_{\text{p}}}^{r_{\text{a}}} \dd r \sqrt{R(r,E,J)} = \Lambda- J\,,
\end{equation}
computed from the solution $p_r^2=R(r,E,J)$ of the energy equation $H(r,p_r,J)=E$, using polar-type coordinates $(r, \varphi)$ in which $J\equiv p_\varphi$ is the angular momentum. The integration is most efficiently achieved by means of the Sommerfeld method, discussed in Ref.~\cite{DJSinv}. Consistently with our previous notation, we introduce the reduced momenta 
\begin{equation}
\hat{\Lambda} = \frac{\Lambda}{G m \mu}\,,\qquad \hat{J} = \frac{J}{G m \mu}\,.
\end{equation}
Inverting $\Lambda=I_r(E,J)+J$, we get $E$ in terms of $\Lambda$ and $J$, which yields the reduced tidal Delaunay Hamiltonian $\hat{H}^\text{D}=H^\text{D}/\mu$, complete up to NNL order: 
\begin{subequations}
\begin{align}
\hat{H}_{\text{pp}}^\text{D} &=- \frac{1}{2 \hat{\Lambda}^2}\nonumber\\
&~+ \frac{1}{c^2} \biggl [- \frac{3}{\hat{J} \hat{\Lambda}^3} + \frac{1}{\hat{\Lambda}^4} \Bigl(\frac{15}{8} -  \frac{1}{8} \nu \Bigr)\biggl]\nonumber\\
&~+ \frac{1}{c^4} \biggl [- \frac{27}{2 \hat{J}^2 \hat{\Lambda}^4}
	+ \frac{1}{\hat{J} \hat{\Lambda}^5} \Bigl(\frac{105}{4}
	- 3 \nu \Bigr)
	+ \frac{1}{\hat{J}^3 \hat{\Lambda}^3} \Bigl(- \frac{35}{4}
	+ \frac{5}{2} \nu \Bigr) + \frac{1}{\hat{\Lambda}^6} \Bigl(- \frac{145}{16} + \frac{15}{16} \nu
	-  \frac{1}{16} \nu^2\Bigr)\biggl]+ \mathcal{O}\left(\frac{1}{c^{6}}\right)\,,\\
%%%%%%%%%%%%%%%%%%%%%%%%%%%%%%%%%%%%%%%%%%%%%%%%%%%%%%%%%%%%%%%%%%%%%
\hat{H}_{\text{tidal}}^\text{D}&=\frac{1}{c^{10}} \biggl(- \frac{9 \tilde{\mu}_{+}^{(2)}}{8 \hat{J}^5 \hat{\Lambda}^7}
	+ \frac{45 \tilde{\mu}_{+}^{(2)}}{4 \hat{J}^7 \hat{\Lambda}^5}
	-  \frac{105 \tilde{\mu}_{+}^{(2)}}{8 \hat{J}^9 \hat{\Lambda}^3} \biggr)\nonumber\\
&~+ \frac{1}{c^{12}} \biggl(- \frac{189 \tilde{\mu}_{+}^{(2)}}{8 \hat{J}^6 \hat{\Lambda}^8}
	+ \frac{675 \tilde{\mu}_{+}^{(2)}}{4 \hat{J}^8 \hat{\Lambda}^6}
	-  \frac{945 \tilde{\mu}_{+}^{(2)}}{8 \hat{J}^{10} \hat{\Lambda}^4}
	+ \frac{1}{\hat{J}^5 \hat{\Lambda}^9} \biggl [\tilde{\mu}_{+}^{(2)} \Bigl(\frac{81}{4}
	-  \frac{27}{8} \nu \Bigr) + \frac{15}{2} \tilde{\sigma}_{+}^{(2)}\biggl]\nonumber\\
	&\qquad\quad
	+ \frac{1}{\hat{J}^9 \hat{\Lambda}^5} \biggl [\tilde{\mu}_{+}^{(2)} \Bigl(\frac{8715}{16}
	- 105 \nu \Bigr)
	+ \frac{525}{16} \tilde{\mu}_{-}^{(2)} \Delta
	+ \frac{945}{2} \tilde{\sigma}_{+}^{(2)}\biggl]\nonumber\\
	&\qquad\quad
	+ \frac{1}{\hat{J}^7 \hat{\Lambda}^7} \biggl [\tilde{\mu}_{+}^{(2)} \Bigl(- \frac{7515}{32}
	+ \frac{765}{16} \nu \Bigr) -  \frac{225}{32} \tilde{\mu}_{-}^{(2)} \Delta
	-  \frac{315}{2} \tilde{\sigma}_{+}^{(2)}\biggl]\nonumber\\
	&\qquad\quad
	+ \frac{1}{\hat{J}^{11} \hat{\Lambda}^3} \biggl [\tilde{\mu}_{+}^{(2)} \Bigl(- \frac{13419}{32}
	+ \frac{945}{16} \nu \Bigr)
	-  \frac{945}{32} \tilde{\mu}_{-}^{(2)} \Delta
	-  \frac{693}{2} \tilde{\sigma}_{+}^{(2)}\biggl]\biggl)\nonumber\\
&~ + \frac{1}{c^{14}} \biggl(\frac{1}{\hat{J}^6 \hat{\Lambda}^{10}} \biggl[\tilde{\mu}_{+}^{(2)}
	\Bigl(\frac{18711}{32} 
	-  \frac{1701}{16} \nu \Bigr)
	+ \frac{405}{2} \tilde{\sigma}_{+}^{(2)}\biggl]\nonumber\\
	&\qquad \quad
	+ \frac{1}{\hat{J}^5 \hat{\Lambda}^{11}} \biggl [\tilde{\mu}_{+}^{(2)} \Bigl(- \frac{31869}{128}
	+ \frac{2349}{32} \nu
	-  \frac{27}{4} \nu^2\Bigr)
	+ \tilde{\sigma}_{+}^{(2)} \Bigl(- \frac{2445}{16}+ \frac{45}{2} \nu \Bigr)\biggl]\nonumber\\
	&\qquad \quad
	+ \frac{1}{\hat{J}^{10} \hat{\Lambda}^6} \biggl [\tilde{\mu}_{+}^{(2)} \Bigl(\frac{285075}{32}
	-  \frac{29025}{16} \nu \Bigr)
	+ \frac{7875}{16} \tilde{\mu}_{-}^{(2)} \Delta
	+ \frac{14175}{2} \tilde{\sigma}_{+}^{(2)}\biggl] \nonumber\\
	&\qquad \quad+ \frac{1}{\hat{J}^{12} \hat{\Lambda}^4} \biggl [\tilde{\mu}_{+}^{(2)} \Bigl(- \frac{32949}{8}
	+ 630 \nu \Bigr)-  \frac{8505}{32} \tilde{\mu}_{-}^{(2)} \Delta
	-  \frac{6237}{2} \tilde{\sigma}_{+}^{(2)}\biggl]\nonumber\\
	&\qquad \quad
	+ \frac{1}{\hat{J}^8 \hat{\Lambda}^8} \biggl [\tilde{\mu}_{+}^{(2)} \Bigl(- \frac{84735}{16}
	+ \frac{9135}{8} \nu \Bigr)
	-  \frac{4725}{32} \tilde{\mu}_{-}^{(2)} \Delta -  \frac{6615}{2} \tilde{\sigma}_{+}^{(2)}\biggl]\nonumber\\
	&\qquad \quad + \frac{1}{\hat{J}^9 \hat{\Lambda}^7} \biggl
	[\tilde{\mu}_{+}^{(2)}\Bigl(- \frac{1082025}{128} 
	+ \frac{129225}{32} \nu
	-  \frac{25515}{64} \nu^2\Bigr)
	+ \tilde{\mu}_{-}^{(2)} \Delta \Bigl(- \frac{139905}{128}
	+ \frac{12075}{64} \nu \Bigr)\nonumber\\
	&\qquad \qquad \quad + \tilde{\sigma}_{+}^{(2)} \Bigl(-12600+ \frac{12285}{4} \nu \Bigr)
	-  \frac{5355}{8} \tilde{\sigma}_{-}^{(2)} \Delta
	-  \frac{1575}{16} \tilde{\mu}_{+}^{(3)}\biggl]\nonumber\\
	&\qquad \quad
	+ \frac{1}{\hat{J}^{13} \hat{\Lambda}^3} \biggl [\tilde{\mu}_{+}^{(2)} \Bigl(- \frac{155331}{16}
	+ \frac{6435}{2} \nu
	-  \frac{10395}{64} \nu^2\Bigr)\nonumber\\
	&\qquad \qquad \quad
	+ \tilde{\mu}_{-}^{(2)} \Delta \Bigl(- \frac{180873}{128}+ \frac{10395}{64} \nu \Bigr)
	+ \tilde{\sigma}_{+}^{(2)} \Bigl(- \frac{188331}{16}
	+ \frac{9009}{4} \nu \Bigr)
	-  \frac{7293}{8} \tilde{\sigma}_{-}^{(2)} \Delta
	-  \frac{3465}{16} \tilde{\mu}_{+}^{(3)}\biggl]\nonumber\\
	&\qquad \quad
	+ \frac{1}{\hat{J}^7 \hat{\Lambda}^9} \biggl [\tilde{\mu}_{+}^{(2)} \Bigl(\frac{645381}{224}
	-  \frac{65325}{56} \nu+ \frac{495}{4} \nu^2\Bigr)
	+ \tilde{\mu}_{-}^{(2)} \Delta \Bigl(\frac{37935}{224}
	-  \frac{225}{8} \nu \Bigr)
	+ \tilde{\sigma}_{+}^{(2)} \Bigl(\frac{25795}{8}
	-  \frac{2625}{4} \nu \Bigr)\nonumber\\
	&\qquad \qquad \quad + \frac{595}{8} \tilde{\sigma}_{-}^{(2)} \Delta
	+ \frac{75}{16} \tilde{\mu}_{+}^{(3)}\biggl]
	\nonumber\\
	&\qquad \quad + \frac{1}{\hat{J}^{11} \hat{\Lambda}^5} \biggl
	[\tilde{\mu}_{+}^{(2)} \Bigl(\frac{928341}{64} 
	-  \frac{95337}{16} \nu
	+ \frac{14175}{32} \nu^2\Bigr)
	+ \tilde{\mu}_{-}^{(2)} \Delta \Bigl(\frac{137943}{64}
	-  \frac{10395}{32} \nu \Bigr)
	\nonumber\\
	&\qquad \qquad \quad + \tilde{\sigma}_{+}^{(2)} \Bigl(\frac{158697}{8}
	-  \frac{18711}{4} \nu \Bigr)+ \frac{11781}{8} \tilde{\sigma}_{-}^{(2)} \Delta
	+ \frac{4725}{16} \tilde{\mu}_{+}^{(3)}\biggl]\biggl)+  \mathcal{O}\left(\frac{\epsilon_\text{tidal}}{c^{6}}\right)\,.
	\end{align}
\end{subequations}
Finally let us provide the expression of the two independent dynamical invariants in the case of quasi-circular orbits. For this purpose, we choose the total conservative energy $E$ and the periastron advance $K$, such that the precession of the orbit in one orbital revolution is $\Delta\Phi = 2\pi K$. These two invariants are expressed in terms of the orbital frequency $\omega = K n$ through the usual invariant post-Newtonian parameter $x=(\frac{Gm\omega}{c^3})^{2/3}$. The energy function has already been computed in~\cite{HFB20a}:
\begin{subequations}\label{eq:Ex}
	\begin{align}
	E_\text{pp} &= -\frac{1}{2} m\nu x c^2 \left[ 1 + \left( -\frac{3}{4}
	-\frac{\nu}{12} \right)x + \left( -\frac{27}{8} +\frac{19}{8}
	\nu - \frac{\nu^{2}}{24} \right)x^{2} \right] +
	\mathcal{O}\left(\frac{1}{c^{6}} \right)\,, \\
	%%%%%%%%%%%%%%%%%%%%%%%%%%%%%%%%%%%%%%%%%%%%%%%%%%%%%%%%%%
	E_\text{tidal} &= -\frac{1}{2} m\nu x c^2
	\biggl\{ - 18 \widetilde{\mu}_{+}^{(2)} x^5 +
	\left[\left(-\frac{121}{2} +33\nu \right)\widetilde{\mu}_{+}^{(2)}-
	\frac{55}{2}\Delta \, \widetilde{\mu}_{-}^{(2)} -
	176\,\widetilde{\sigma}_{+}^{(2)} \right]x^6 \nonumber\\ 
	&\qquad\quad + \left[ \left(-\frac{20865}{56} +
	\frac{5434}{21}\nu -\frac{91}{4}\nu^2 \right)
	\widetilde{\mu}_{+}^{(2)} +
	\Delta \left(-\frac{11583}{56}+\frac{715}{12}\nu \right)
	\widetilde{\mu}_{-}^{(2)} \right. \nonumber\\& \qquad\qquad\qquad \left. +
	\left(-\frac{2444}{3} +
	\frac{1768}{3}\nu \right)\widetilde{\sigma}_{+}^{(2)} -
	\frac{884}{3} \Delta \, \widetilde{\sigma}_{-}^{(2)} -
	130 \,\widetilde{\mu}_{+}^{(3)} \right]x^7 \biggr\} + \mathcal{O}\left(
	\frac{\epsilon_\text{tidal}}{c^{6}} \right)\,,
	\end{align}
\end{subequations}
while the orbital precession (for circular orbits) is given by
\begin{subequations}
\begin{align}
K_{\text{pp}}={}& 1 + 3 x
+ \Bigl(\frac{27}{2}
- 7 \nu \Bigr) x^2 +
\mathcal{O}\left(\frac{1}{c^{6}} \right)\,,\\
%%%%%%%%%%%%%%%%%%%%%%%%%%%%%%%%%%%%%%%%%%%%%%%%%%%%%%
K_{\text{tidal}}={}&45 \tilde{\mu}_{+}^{(2)} x^5
+ \biggl [\tilde{\mu}_{+}^{(2)} \Bigl(\frac{1755}{4}
- 120 \nu \Bigr)
+ \frac{315}{4} \tilde{\mu}_{-}^{(2)} \Delta
+ 624 \tilde{\sigma}_{+}^{(2)}\biggl] x^6
+ \biggl [\tilde{\mu}_{+}^{(2)} \Bigl(\frac{64911}{16}
-  \frac{9381}{4} \nu
+ 140 \nu^2\Bigr)\nonumber\\
& + \tilde{\mu}_{-}^{(2)} \Delta \Bigl(\frac{19191}{16}
-  \frac{945}{4} \nu \Bigr)
+ \tilde{\sigma}_{+}^{(2)} \Bigl(6220
- 2544 \nu \Bigr)
+ 1156 \tilde{\sigma}_{-}^{(2)} \Delta
+ 420 \tilde{\mu}_{+}^{(3)}\biggl] x^7 + \mathcal{O}\left(
\frac{\epsilon_\text{tidal}}{c^{6}} \right)\,.
\end{align}
\end{subequations}
The point-particle part of the periastron advance is displayed at the 2PN order so as to be consistent with the NNL tidal part (but see Refs.~\cite{DJS15eob, BBBFMb} for the expression at 4PN order).

\section{Summary}

Starting from the effective Lagrangian for tidal interactions between compact objects investigated in our previous work~\cite{HFB20a}, we derived the corresponding tidal Hamiltonian in ADM-like and isotropic coordinates up to NNL/7PN order $\mathcal{O}(\epsilon_\text{tidal}/c^4)$, \textit{i.e.}, formally $(v/c)^{14}$. We then checked that our result agrees with the recent literature from the amplitude/EFT community~\cite{Cheung:2020sdj,Kalin:2020lmz} up to the NL/6PN order for all the NL PM terms $G^2$ and $G^3$. However, some additional terms $G^4$ arise to NNL/7PN order for the mass and current quadrupoles and are consistently included. Finally, we provided the tidal Delaunay Hamiltonian as well as the orbital precession for circular orbits up to NNL/7PN order.

\acknowledgments

We thank Mikhail Solon for a discussion that prompted the computation of the tidal post-Newtonian Hamiltonian presented in this paper. 

%\appendix

\bibliography{ListeRef_HFB20c}

\end{document}